\documentclass{aastex631} 

\begin{document}

\title{Eighteen Exoplanet Host Stars from the NPOI Data Archive}

\author{Ellyn K. Baines}
\affil{Naval Research Laboratory, Remote Sensing Division, 4555 Overlook Ave SW, Washington, DC 20375, USA}
\email{ellyn.k.baines.civ@us.navy.mil}

\author{Jeremy Jones}
\affil{Center for High Angular Resolution Astronomy and Department of Physics and Astronomy, Georgia State University, 25 Park Place, Suite 605, Atlanta, GA 30303, USA}

\author{James H. Clark III}
\affil{Naval Research Laboratory, Remote Sensing Division, 4555 Overlook Ave SW, Washington, DC 20375, USA}

\author{Henrique R. Schmitt}
\affil{Naval Research Laboratory, Remote Sensing Division, 4555 Overlook Ave SW, Washington, DC 20375, USA}

\author{Jordan M. Stone}
\affil{Naval Research Laboratory, Remote Sensing Division, 4555 Overlook Ave SW, Washington, DC 20375, USA}

\begin{abstract}

During the course of publishing angular diameters from the Navy Precision Optical Interferometer data archive, we found we had data on 17 confirmed exoplanet host stars and one exoplanet candidate (HD 20902/$\alpha$ Per). Here, we update our previously published stellar radii with more precise $Gaia$ parallaxes when available, and use our radius and effective temperature measurements to fit each star's mass and age using MIST models. The mass changed by more than 10$\%$ for 9 of the 18 stars. Combining our updated masses, radii, and temperatures, we present refined planetary masses as well as habitable zone calculations. 

\end{abstract}

\keywords{stars: fundamental parameters, techniques: high angular resolution, techniques: interferometric}


\section{Introduction} 

The need for high-precision stellar diameters grows, especially in their capacity to act as benchmarks for large stellar surveys such as \emph{Gaia} \citep{2020MNRAS.493.2377R}. Precise diameters are also vital when it comes to characterizing exoplanets where we only know the planet as well as we know the host star. Optical/infrared interferometry has made important contributions to exoplanet research, including directly measuring the size of a transit-planet host and therefore directly obtaining the size of the planet \citep{2007ApJ...661L.195B, 2015MNRAS.447..846B, 2019AandA...631A..92L, 2021AJ....162..118E}, efforts to image the starspots on an exoplanet host \citep{2022AJ....163...19R}, investigating the inclination of the nearby solar analog HD 10700/$\tau$ Cet and what the nearly pole-on orientation means for its exoplanets \citep{2023AJ....166..123K}, confirming the single nature of the exoplanet host star HD 179070/Kepler-21 \citep{2012MNRAS.423L..16H}, and determining the extent of the habitable zone of a cool star \citep{2011ApJ...729L..26V}. Interferometry has also been used to characterize the orbit of an unusual planet that particularly challenges formation theories \citep[HIP 65426 b,][]{2023AJ....166..257B}, imaging and discovering planets \citep{2020AandA...642L...2N, 2023AandA...671L...5H}, and characterizing planetary atmospheres \citep{2019AandA...623L..11G, 2021AJ....161..148W, 2024arXiv240403776N}. 

\begin{figure}[h]
\includegraphics[width=0.85\textwidth]{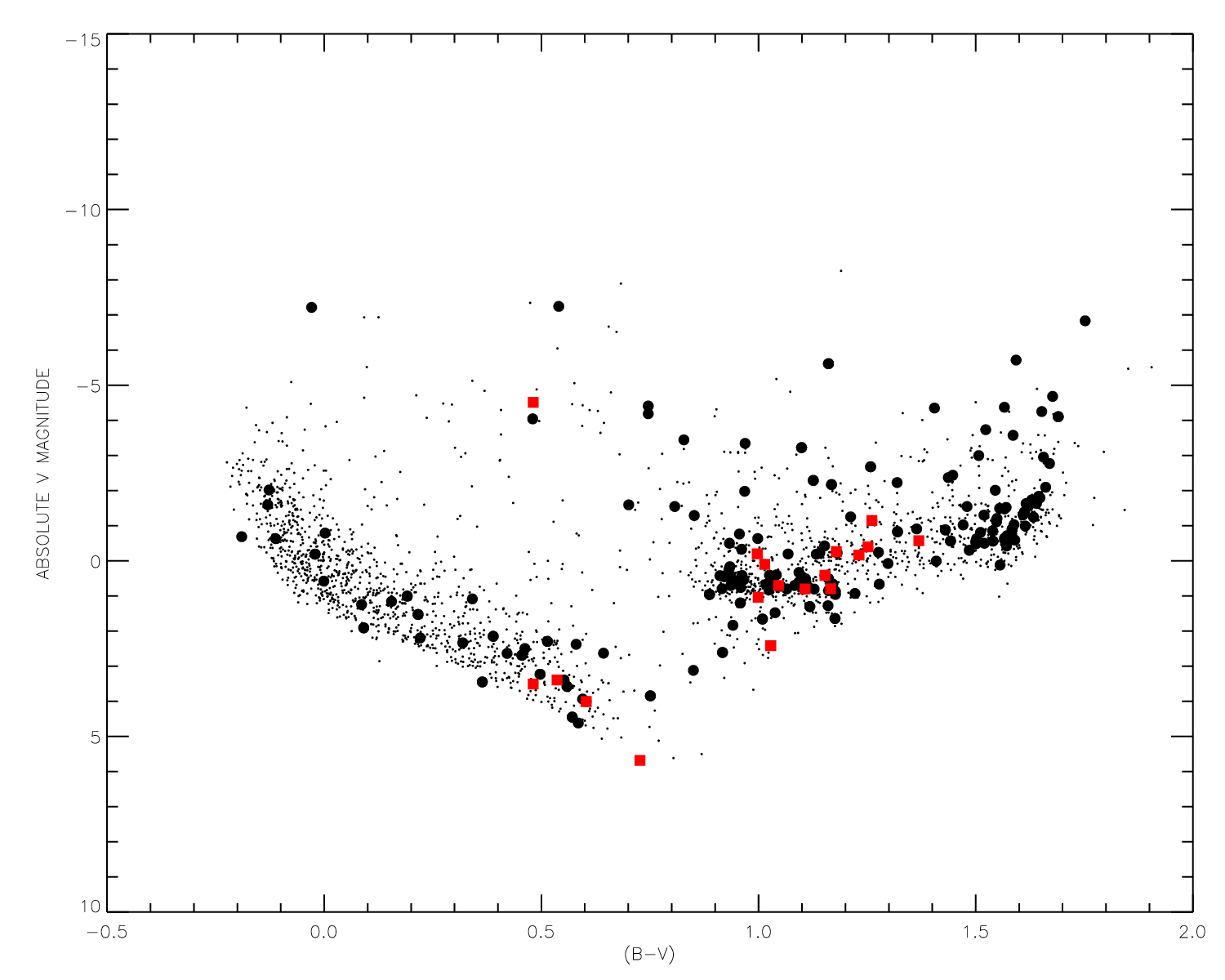}
\caption{A color-magnitude diagram of the exoplanet host stars presented here (red squares), past NPOI targets (large black circles), and targets from \emph{JSDC} \citep[][small black points]{2014ASPC..485..223B} that fall within the limits of the NPOI observable range of declination north of -10 deg and brighter than $V=6.0$.}
  \label{color_mag}
\end{figure}

As part of our efforts to publish angular diameters in the Navy Precision Optical Interferometer (NPOI) data archive \citep{2018AJ....155...30B, 2021AJ....162..198B, 2023AJ....165...41B}, we discovered we had data on 18 exoplanet host stars, one of them with an unconfirmed planet as of yet (HD 20902/$\alpha$ Per). Table \ref{sample} lists each star's identifiers, $V$ magnitude, parallax, limb-darkened angular diameter ($\theta_{\rm LD}$), physical radius ($R$), and effective temperature ($T_{\rm eff}$). Figure \ref{color_mag} shows a color-magnitude diagram that includes the 18 exoplanet host stars studied here, previously published NPOI diameters, and targets from the \emph{JMMC Stellar Diameter Catalog} \citep[$JSDC$,][]{2014ASPC..485..223B} that fall within the NPOI's observable parameters of brighter than 6 mag in the $V$-band and north of -10 deg in declination. 

The paper is organized as follows: Section 2 considers the stars and their parallaxes, Section 3 details our process of fitting stellar ages and masses to our $R$ and $T_{\rm eff}$ measurements, Section 4 synthesizes that information to determine planetary masses and the habitable zone around each star, and Section 5 is the summary and discussion. 

\vspace{0.25in}

\section{Updating Stellar Radii}

As we explored the NPOI's data archive, we amassed a collection of 178 stars covering a wide range of spectral and luminosity types. From those, we built a sample of 18 exoplanet host stars that were detected using the radial velocity method. We set out to see what we could learn about the host stars and their planets using the $R$ and $T_{\rm eff}$ measurements from NPOI data.

In the original publications referenced above, some of the exoplanet host stars used older \emph{Hipparcos} parallax measurements \citep{2007AandA...474..653V} because that was what was available at the time, so we began by updating the parallaxes from \emph{Gaia} \citep{2022yCat.1355....0G} when applicable. Table \ref{sample} lists each star's radius as it was published as well as the updated radius using the $Gaia$ parallax. The $Gaia$ Data Release Notes indicate a full astrometric solution for stars up to a brightness of $G \approx 3$ mag, and three of our stars are around that cut-off: HD 137759/$\iota$ Dra $G = 2.97$ mag, HD 163917/$\nu$ Oph $G = 3.08$ mag, and HD 222404/$\gamma$ Cep $G = 2.94$ mag. We compared the $Gaia$ parallax to the $Hipparcos$ parallax for each, and found a percent difference between them of 1$\%$, 6$\%$, and 2$\%$, respectively. We will optimistically use the $Gaia$ parallaxes here, considering the stars are not significantly brighter than $G = 3$ mag.

\begin{figure}[h]
\includegraphics[width=1.0\textwidth]{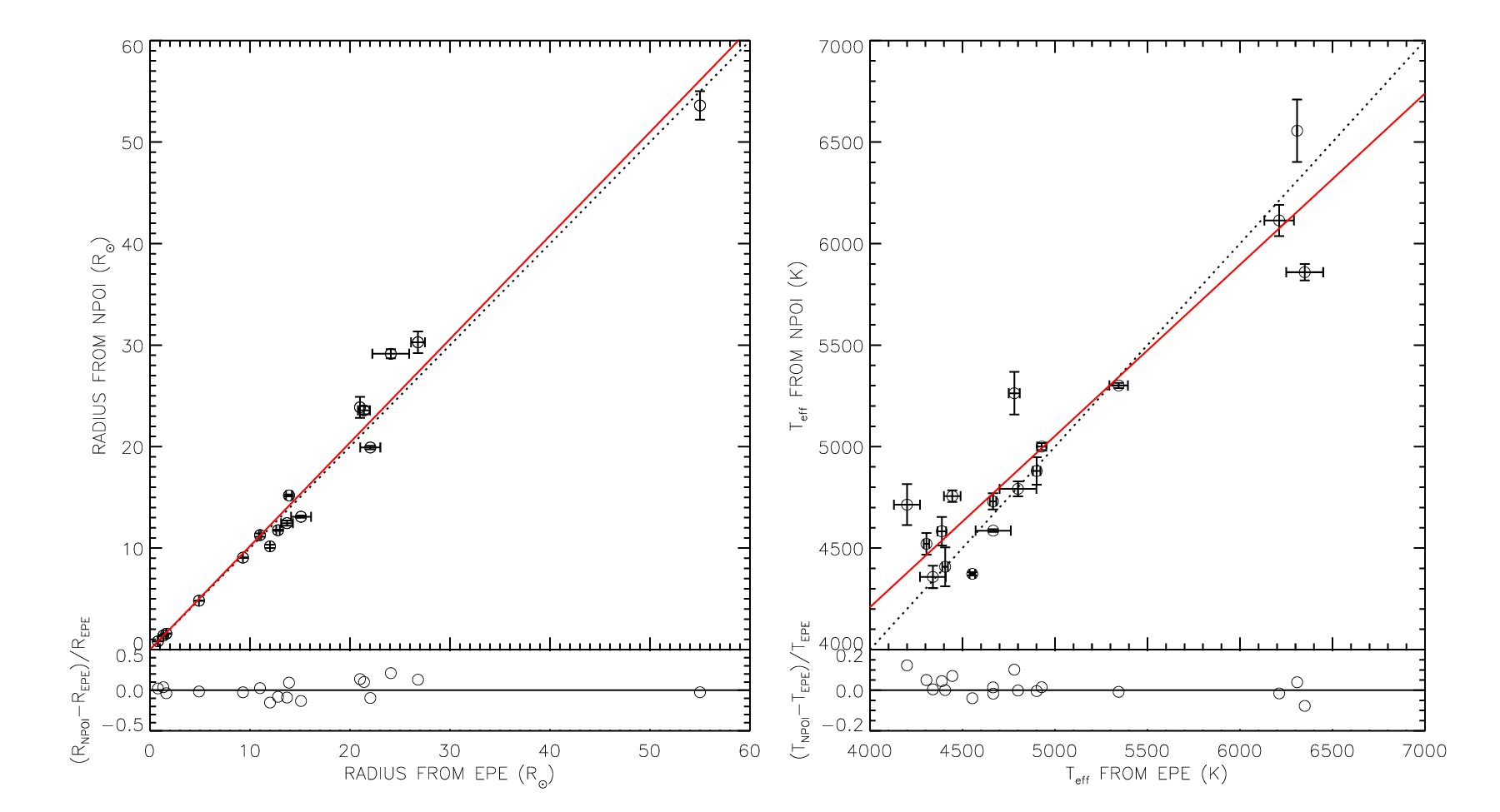}
\caption{Comparing the stellar radius (\emph{left panel}) and effective temperature (\emph{right panel}) from \emph{EPE} with those determined from NPOI measurements. The dotted lines show the 1:1 ratio, and the solid red lines are the linear fits of $f(x) = 1.020 x - 0.003$ and $f(x) = 0.845 x - 834.34$ for the radius and effective temperature, respectively. The bottom parts of each plot show the residuals to the 1:1 ratio line.}
  \label{rad_teff}
\end{figure}

We also include $R$ and $T_{\rm eff}$, when available, from \emph{The Encyclopaedia of Exoplanetary Systems} (\emph{EPE}, exoplanet.eu) as a comparison to NPOI values. These are listed in Table \ref{sample}. Some values did not have errors indicated so a 5$\%$ error was assigned for the radius for HD 20902/$\alpha$ Per, HD 62509/$\beta$ Gem, HD 143107/$\epsilon$ CrB, HD 188310/$\xi$ Aql, and HD 222404/$\gamma$ Cep, and a 2$\%$ error was added for $T_{\rm eff}$ for HD 120136/$\tau$ Boo. Figure \ref{rad_teff} shows how the $R$ and $T_{\rm eff}$ from the NPOI and \emph{EPE} compare. The radii have a consistent linear fit\footnote{The linear fit used here is a simple one, fitting the data with the linear model $y = A + Bx$ and minimizing the $\chi^{\rm 2}$ error statistic. This acts as a generalization of how much the data differ between sources.} of $f(x) = 1.020 x - 0.003$, while the temperature is more offset at $f(x) = 0.845 x - 834.34$.

\section{Calculating Age and Mass}

For each star in the sample, we calculated a probability distribution function (PDF) for both the age and the mass by comparing the derived luminosity ($L$) and T$_\mathrm{eff}$ with those predicted by the MESA \citep[Modules for Experiments in Stellar Astrophysics,][]{2011ApJS..192....3P, 2013ApJS..208....4P, 2015ApJS..220...15P, 2018ApJS..234...34P} evolution models, using the MIST \citep[MESA Isochrones \& Stellar Tracks,][]{2016ApJS..222....8D, 2016ApJ...823..102C} web interpolator. 

The core product generated by evolution models is the stellar mass tracks. Isochrones are generated based on these mass tracks. They are, in essence, both representations of the models that built them. The key difference between the two is what they have as their independent variable: i.e., initial mass in the case of mass tracks and age in the case of isochrones. The rationale for using mass tracks to determine age is that the mass tracks provide much higher resolution in age than we could achieve from even a large set of isochrones. The inverse is true for using isochrones to determine the mass. Because the isochrones and the mass tracks are both representations of the same underlying model, they are complementary to each other in this task \citep[see][which includes a good discussion on the realationship between mass tracks and isochrones]{2016ApJS..222....8D}.

\subsection{Age Calculation}
\label{sec:age_calc}

For each star, we downloaded a grid of mass tracks at the star's published metallicity as listed in Table \ref{sample}. Because of the large variety of stars in our sample, the range of masses in this grid varied from star to star. In each case, we ran the following procedure iteratively, changing the step size or the range of the mass tracks we downloaded as needed until the resulting age distribution did not change. In practice, this requires having enough mass tracks to cover the luminosity and temperature of $\sim 4 \times$ the uncertainty in the observations.

Once we had our grid of mass tracks, we interpolated along the track to generate a new mass track with the same shape, but with uniform age increments of 10 kyr. At each point along each mass track, we calculated a gaussian weight based on its distance to the measured value of $L$ and $T_\mathrm{eff}$. This weight takes the form $w = e^{-\sigma^2/2}$ where $\sigma = \sqrt{((L_m-L_o)/L_{\rm err})^2+((T_m-T_o)/T_{\rm err})^2}$. The $X_m$ subscript indicates the modeled value, the $X_o$ subscript indicates the observed value, and the $X_{\rm err}$ subscript indicates the uncertainty in the observed value. The PDF of the age is the distribution of all points along the interpolated mass track where each point was weighted by the gaussian weight. This allowed us to account for regions of the HR diagram where $L$ and $T_\mathrm{eff}$ evolve at different rates over time and regions where there is overlap of different periods of evolution (e.g., sub giant branch, red giant branch, blue loop, etc.). 

One example of where this is particularly useful is $\upsilon$ And (see Figure 3), which has $L$ and $T_\mathrm{eff}$ such that its uncertainties overlap mass tracks in both the red giant branch phase and the core helium burning phase. By giving a weight to all values on the mass track, we can better account for each of these phases, as stars will spend much longer in the core helium burning phase than in the red giant phase, but $\upsilon$ And's $L$ and $T_\mathrm{eff}$ is centered in the red giant phase. This being the case, the points on the mass track near the center of the $L$ and $T_\mathrm{eff}$ measurements are weighted higher, but there are more of them (each at a lower weight) in the core helium burning phase. This allows us to appropriately account for the likelihood of the age. In general, we adopted an age for each star that is the median of the weighted distribution, and calculated confidence intervals (CIs) that correspond to 1- and 2-$\sigma$ uncertainties in the age (see Table \ref{ages} and Figure 3).

One of the stars in the presented sample (HD 120136/$\tau$ Boo) has an uncertainty in its $T_{\rm eff}$ that overlaps the zero age main sequence. Therefore cannot calculate a median age and only present the upper bounds of its 1- and 2-$\sigma$ CIs.

\subsection{Mass Calculation}
\label{sec:mass_calc}

For calculating the masses, we followed an analogous procedure as that presented in Section \ref{sec:age_calc}. The key difference is that we used isochrones instead of mass tracks, and we interpolated along each isochrone to generate a new isochrone with the same shape, but with uniform mass increments of 0.0001 M$_\odot$. Otherwise the procedure for calculating the PDF of the mass was the same as that for the age (see Table \ref{mass_table} and Figure \ref{star_mass}). Based on the PDF we calculated from this procedure for each star, we adopted a mass that is the median of the weighted distribution, and calculated CIs that correspond to 1- and 2-$\sigma$ uncertainties in the mass.

As shown in Figure \ref{star_mass}, we see that the mass errors resulting from our fits are sometimes smaller than those presented in \emph{EPE}, and there is some offset between the masses themselves. One reason for this may be the fact that we used a consistent methodology based on one observing technique, while \emph{EPE} presents masses determined across a range of techniques. For example, in \emph{EPE}, HD 9826/$\upsilon$ And's mass was determined using evolutionary tracks \citep{1998AandA...336..942F}, HD 10700/$\tau$ Cet's mass used solar-type oscillation measurements (\emph{EPE} cites \citet{2013AandA...551A..79T} as the source of the mass, though that paper in turn cites \citet{2009AandA...494..237T}), HD 20902/$\alpha$ Per's mass used evolutionary tracks based on $T_{\rm eff}$ and surface gravity (\emph{EPE} cites the preprint that corresponds to \citet{2012AandA...543A..37L}, which in turn cites \citet{2010MNRAS.402.1369L} for the mass), and HD 62509/$\beta$ Gem's mass used p-mode oscillations \citep{2012AandA...543A..98H}. The \emph{EPE} sources also often use older data, which sometimes lacks the improved precision of newer data.

\section{Planetary Masses and Habitable Zones}

Table \ref{planet_sys} lists the planetary system parameters from \emph{EPE}, and when asymmetric error bars were indicated in \emph{EPE}, the larger of the two was used here. We used our best fit mass and the parameters in Table \ref{planet_sys} to calculate the mass function 
\begin{equation}
f(m) = \frac{(m_{\rm p} \sin i)^3}{(M_{\star} + m_{\rm p})^2} = \frac{P}{2 \pi G} (K \sqrt{1-e^2})^3,
\end{equation}
where $m_{\rm p}$ is the planetary mass, $i$ is inclination, $M_\star$ is the stellar mass, $P$ is the planet's orbital period, $K$ is the velocity semiamplitude in m s$^{\rm -1}$, $e$ is eccentricity, and $G$ is the gravitational constant. 

We calculated planetary masses by setting $f(m)$ equal to the far right side of Equation 1, determining $f(m)$, and finally solving for $m_{\rm p} \sin i$ by assuming that the planetary mass was negligible compared to the stellar mass. That was our first iteration. We then folded $m_{\rm p} \sin i$ into the calculation, which changed $m_{\rm p}$ by a small amount, an average of 0.2$\%$ and a maximum of 0.5$\%$. All masses converged in two iterations, and Table \ref{planet_mass_table} shows the results (see also Figure \ref{planet_mass}). 

When determining the error on $m_{\rm p} \sin i$, we used the larger of the 1-$\sigma$ errors if the stellar mass had asymmetric error bars, and we did not include the error from $e$. If we had done so, the errors on $m_{\rm p} \sin i$ would have been worthlessly large, considering that the uncertainties on $e$ are generally poorly constrained.

Another important parameter for characterizing exoplanets is whether or not they lie within the habitable zone of the their parent star, the region where liquid water can exist on the planet's surface. We used the equations from \citet{2012PASP..124..323K} to calculate the extent of each star's habitable zone using our $T_{\rm eff}$ measurements:
\begin{equation}
S_{i} = 4.190 \times 10^{-8} \; T_{\rm eff}^2 - 2.139 \times 10^{-4} \; T_{\rm eff} + 1.268,
\end{equation}
and 
\begin{equation}
S_{o} = 6.190 \times 10^{-9} \; T_{\rm eff}^2 - 1.319 \times 10^{-5} \; T_{\rm eff} + 0.2341,
\end{equation}
where $S_{i}$ and $S_{o}$ are the stellar fluxes at the inner and outer boundaries of the habitable zone in units of the solar constant. The inner and outer physical boundaries $r_{i,o}$ in AU were then calculated using
\begin{equation}
r_i = \sqrt{ \frac{L/L_\odot}{S_{i}} } \; \; \; \; \; {\rm and} \; \; \; \; \; r_o = \sqrt{ \frac{L/L_\odot}{S_{o}} },
\end{equation}
and Table \ref{hab_zone} shows the results of these calculations.

Only one star here has $e=0$ (HD 188310/$\xi$ Aql), and the rest exhibit a range from $e =$ 0.2 to 0.7 (see Table \ref{planet_sys}). This naturally has an impact on the planet's ability to sustain liquid water on its surface \citep[see, e.g.,][]{2014PNAS..11112641K, 2023ApJ...943L...1J}. Comparing the HZ values calculated with planetary separations from \emph{EPE}, we find that only HD 9826/$\upsilon$ And d lies within the habitable zone for its host star for its entire orbit, which is not a surprise \citep[see, e.g.,][]{2021AJ....162...30S}. HD 10700/$\tau$ Cet f could spend some time in the HZ \citep[see][who do a deep dive into the $\tau$ Cet system]{2021AJ....161...17D}. The rest of the planets orbit inside the inner boundary of the HZ.

Another aspect to consider is the evolved state of most of the exoplanet hosts in the sample. The HZ evolves over time with the star itself due to many reasons, including how the stellar luminosity varies. \citet{2014AandA...567A.133V} consider this in detail for low-mass stars (0.70 - 1.10 $M_\odot$), and found that stellar metallicity and mass affect HZ calculations, especially for the inner HZ. More precisely measured stellar masses can help with these determinations. 

\section{Summary and Discussion}

We used interferometric radii and effective temperatures to determine exoplanet host stars' ages and masses, and used the updated stellar masses to calculate each star's habitable zone and the mass(es) of the planet(s) in the system. The majority of exoplanet host stars tend to be main-sequence stars, which is a natural result of the radial velocity and transit detection techniques. Nevertheless, some studies have searched for exoplanets around evolved stars \citep[see, e.g.,][]{2005AandA...437..743H, 2015AandA...574A.116R, 2015AandA...573A..36N}. While these programs and observations from $TESS$ have increased the number of evolved stars known to host planets, the total is still a small percentage when compared to dwarf stars: on the order of $\sim$100 versus the $\sim$5,000 candidates for main-sequence hosts \citep{2023AJ....165...44G}. Because many of the stars presented here are evolved, work like this can act as an interesting link between planets around main-sequence stars and those found orbiting white dwarfs, as in the recent results from \citet{2024ApJ...962L..32M}, who used $JWST$ to directly image giant planets around nearby white dwarfs. Characterizing the stars and planets of evolved systems can help to constrain our picture of how planetary orbits evolve with their stars, and models of planet engulfment \citep{2022AandA...661A..63W}. 

\begin{acknowledgements}

This material was based upon work supported by the National Aeronautics and Space Administration under Grant 18-XRP18$\_$2-0017 issued through the Exoplanets Research Program. The Navy Precision Optical Interferometer is a joint project of the Naval Research Laboratory and the U.S. Naval Observatory, and is funded by the Office of Naval Research and the Oceanographer of the Navy. This research has made use of the SIMBAD database and Vizier catalogue access tool, operated at CDS, Strasbourg, France. Finally, this research has made extensive use of data obtained from or tools provided by the portal exoplanet.eu of \emph{The Extrasolar Planets Encyclopaedia}.

\end{acknowledgements}

\clearpage


\startlongtable
\begin{longrotatetable}
\begin{deluxetable}{cccccccccccccccc}
\tablewidth{0pc}
\tabletypesize{\scriptsize}
\tablecaption{Sample Stars.\label{sample}}
\tablehead{
 \colhead{} & \colhead{} & \colhead{} & \colhead{Other} & \colhead{Spectral} & \colhead{$V$} & \colhead{Parallax} & \colhead{} & \colhead{} & \colhead{$\theta_{\rm LD}$} & \colhead{ } & \colhead{$R_{\rm published}$} & \colhead{$R_{\rm updated}$} & \colhead{$T_{\rm eff}$} & \colhead{$R_{\rm EPE}$} & \colhead{$T_{\rm EPE}$} \\
 \colhead{HD} & \colhead{FK5} & \colhead{HR} & \colhead{Name}  & \colhead{Type} & \colhead{(mag)} & \colhead{(mas)} & \colhead{Ref} & \colhead{[Fe/H]} & \colhead{(mas)} & \colhead{Source} & \colhead{($R_\odot$)} & \colhead{($R_\odot$)} & \colhead{(K)} & \colhead{($R_\odot$)} & \colhead{(K)} \\ }
\startdata
9826   & 1045 & 458  & $\upsilon$ And & F9 V      & 4.09 & 74.194$\pm$0.208 & 1 &  0.08 & 1.083$\pm$0.018 & 3 & 1.57$\pm$0.03  & 1.57$\pm$0.03  & 6114$\pm$77 & 1.631$\pm$0.014 & 6212$\pm$80 \\
10700  & 59   & 509  & $\tau$ Cet     & G8 V      & 3.50 & 273.96$\pm$0.17  & 2 & -0.51 & 2.072$\pm$0.010 & 4 & 0.81$\pm$0.01  & 0.81$\pm$0.01  & 5301$\pm$13 & 0.793$\pm$0.004 & 5344$\pm$50 \\
12929  & 74   & 617  & $\alpha$ Ari   & K2 III    & 2.01 & 49.56$\pm$0.25   & 2 & -0.17 & 7.006$\pm$0.027 & 5 & 15.19$\pm$0.10 & 15.19$\pm$0.10 & 4373$\pm$8 & 13.9$\pm$0.3 & 4553$\pm$15 \\
20902  & 120  & 1017 & $\alpha$ Per   & F5 I      & 1.80 & 6.44$\pm$0.17    & 2 &  0.14 & 3.180$\pm$0.008 & 3 & 53.07$\pm$$^{\rm 1.37}_{\rm 1.45}$ & 53.07$\pm$1.41 & 5859$\pm$41 & 55.0 & 6350$\pm$100 \\
28305  & 164  & 1409 & $\epsilon$ Tau & G9.5 III  & 3.54 & 22.365$\pm$0.172 & 1 &  0.14 & 2.592$\pm$0.050 & 6 & 12.53$\pm$0.28 & 12.46$\pm$0.26 & 4880$\pm$67 & 13.7$\pm$0.6 & 4901$\pm$20 \\
54719  & -    & 2697 & $\tau$ Gem     & K2 III    & 4.41 & 8.326$\pm$0.159  & 1 &  0.13 & 2.346$\pm$0.069 & 5 & 30.27$\pm$$^{\rm 1.08}_{\rm 1.09}$ & 30.28$\pm$1.06 & 4583$\pm$68 & 26.8$\pm$0.7 & 4388$\pm$25 \\
62509  & 295  & 2990 & $\beta$ Gem    & K0 III    & 1.14 & 96.54$\pm$0.27   & 2 &  0.08 & 8.134$\pm$0.013 & 6 & 9.06$\pm$0.03  & 9.06$\pm$0.03   & 4586$\pm$7 & 9.3 & 4666$\pm$95 \\
66141  & 2623 & 3145 & 13 Pup         & K2 III    & 4.39 & 12.523$\pm$0.114 & 1 & -0.26 & 2.747$\pm$0.039 & 6 & 22.99$\pm$$^{\rm 0.55}_{\rm 0.56}$ & 23.57$\pm$0.40 & 4521$\pm$53 & 21.4$\pm$0.6 & 4305$\pm$15 \\
120136 & 507  & 5185 & $\tau$ Boo     & F7 IV-V   & 4.50 & 64.047$\pm$0.109 & 1 &  0.24 & 0.822$\pm$0.049 & 6 & 1.38$\pm$0.08  & 1.38$\pm$0.08 & 6556$\pm$197 & 1.331$\pm$0.027 & 6309 \\
136726 & -    & 5714 & 11 UMi         & K4 III    & 5.01 & 7.926$\pm$0.087  & 1 & 0.05 & 2.149$\pm$0.023 & 6 & 28.20$\pm$$^{\rm 0.71}_{\rm 0.73}$ & 29.14$\pm$0.45 & 4358$\pm$55 & 24.08$\pm$1.84 & 4340$\pm$70 \\
137759 & 571  & 5744 & $\iota$ Dra    & K2 III    & 3.29 & 32.524$\pm$0.131 & 1 &  0.09 & 3.559$\pm$0.011 & 6 & 11.87$\pm$0.05 & 11.76$\pm$0.06 & 4756$\pm$29 & 12.8$\pm$0.17 & 4445$\pm$45 \\
141004 & -    & 5868 & $\lambda$ Ser  & G0 V      & 4.43 & 83.921$\pm$0.150 & 1 & -0.02 & 0.982$\pm$0.056 & 5 & 1.26$\pm$0.07 & 1.26$\pm$0.07 & 6087$\pm$174 & N/A & N/A \\
143107 & 593  & 5947 & $\epsilon$ CrB & K2 III    & 4.14 & 13.492$\pm$0.102 & 1 & -0.20 & 2.997$\pm$0.128 & 6 & 21.87$\pm$$^{\rm 0.98}_{\rm 0.99}$ & 23.87$\pm$1.04 & 4408$\pm$96 & 21.0 & 4406$\pm$15 \\
163917 & 673  & 6698 & $\nu$ Oph      & G9 III    & 3.34 & 22.903$\pm$0.192 & 1 &  0.12 & 2.789$\pm$0.005 & 6 & 13.85$\pm$0.17 & 13.09$\pm$0.11 & 5000$\pm$17 & 15.1$\pm$1.0 & 4928$\pm$25 \\
170693 & 3465 & 6945 & 42 Dra         & K1.5 III  & 4.83 & 11.056$\pm$0.084 & 1 & -0.49 & 2.048$\pm$0.009 & 6 & 21.25$\pm$$^{\rm 0.41}_{\rm 0.43}$ & 19.91$\pm$0.17 & 4714$\pm$101 & 22.03$\pm$1.00 & 4200$\pm$70 \\
188310 & -    & 7595 & $\xi$ Aql      & G9.5 III  & 4.70 & 17.518$\pm$0.101 & 1 & -0.23 & 1.658$\pm$0.025 & 6 & 10.03$\pm$0.22 & 10.17$\pm$0.16 & 5263$\pm$105 & 12.0 & 4780$\pm$30 \\
219449 & 1608 & 8841 & 91 Aqr         & K1 III    & 4.23 & 22.092$\pm$0.130 & 1 &  0.00 & 2.220$\pm$0.031 & 6 & 10.96$\pm$0.21 & 10.80$\pm$0.16 & 4730$\pm$40 & 11.0$\pm$0.1 & 4665$\pm$18 \\
222404 & 893  & 8974 & $\gamma$ Cep   & K1 III-IV & 3.21 & 72.517$\pm$0.147 & 1 &  0.15 & 3.254$\pm$0.020 & 6 & 4.93$\pm$0.04  & 4.82$\pm$0.03 & 4792$\pm$37 & 4.9 & 4800$\pm$100 \\
\enddata
\tablecomments{HD 20902/$\alpha$ Per hosts an unconfirmed planet, but we treat it here as if it is genuine. \\
Spectral types are from SIMBAD; $V$ magnitudes are from \citet{Mermilliod}; parallaxes are from (1) $Gaia$ $DR3$ \citep{2022yCat.1355....0G} and (2) \citet{2007AandA...474..653V}; [Fe/H] is from \citet{2012AstL...38..331A} except for HD 10700, which is from \citet{2016ApJ...826..171G}; the limb-darkened diameter $\theta_{\rm LD}$ is from the source listed in the next column, which are: (3) \citet{2021AJ....162..198B}, (4) \citet{2014ApJ...781...90B}, (5) \citet{2023AJ....165...41B}, and (6) \citet{2018AJ....155...30B}; $R_{\rm published}$ is the diameter from the original publication; and $R_{\rm updated}$ is from combining $\theta_{\rm LD}$ with $Gaia$ parallaxes that were not available at the time of previous publications. The columns subscripted \emph{EPE} are from \emph{The Encyclopaedia of Exoplanetary Systems}.}
\end{deluxetable}
\end{longrotatetable}

\clearpage


\begin{deluxetable}{ccll}
\tablewidth{0pc}
\tabletypesize{\normalsize}
\tablecaption{Stellar Ages from MIST Models.\label{ages}}
\tablehead{\colhead{} & \colhead{Spectral} & \colhead{Age,$1\sigma$} & \colhead{Age,$2\sigma$} \\
\colhead{HD} & \colhead {Type} & \colhead{(Myr)} & \colhead{(Myr)}}
\startdata
9826	& F9 V      & 4022$^{\rm +1123}_{\rm -394}$  &  4022$^{\rm +1915}_{\rm -817}$  \\			
10700	& G8 V      & 4099$^{\rm 1259+}_{\rm -1239}$ &  4099$^{\rm +2563}_{\rm -2435}$   \\ 
12929	& K2 III    & 9192$^{\rm +337}_{\rm -616}$   &  9192$^{\rm +1042}_{\rm -899}$  \\
20902	& F5 I      & 69$^{\rm +4}_{\rm -2}$         &  69$^{\rm +6}_{\rm -4}$  \\
28305	& G9.5 III  & 580$^{\rm +119}_{\rm -93}$     &  580$^{\rm +239}_{\rm -125}$   \\
54719	& K2 III    & 211$^{\rm +33}_{\rm -24}$      &  211$^{\rm +69}_{\rm -65}$   \\
62509	& K0 III    & 3844$^{\rm +204}_{\rm -97}$    &  3844$^{\rm +141}_{\rm -283}$  \\
66141	& K2 III    & 1098$^{\rm +254}_{\rm -274}$   &  1098$^{\rm +567}_{\rm -522}$   \\
120136	& F6 IV     & $<$2044 & $<$3001   \\
136726	& K4 III    & 639$^{\rm +259}_{\rm -143}$    &  639$^{\rm +507}_{\rm -278}$  \\
137759	& K2 III    & 1449$^{\rm +79}_{\rm -216}$    &  1449$^{\rm +309}_{\rm -713}$  \\ 
141004	& G0 V      & 4534$^{\rm +2387}_{\rm -2449}$ &  4534$^{\rm +4945}_{\rm -3906}$ \\
143107	& K2 III    & 1600$^{\rm +1420}_{\rm -447}$  &  1600$^{\rm +3610}_{\rm -911}$ \\
163917	& G9 III    & 344$\pm$6                      &  344$\pm$9   \\
170693	& K1.5 III  & 1129$^{\rm +629}_{\rm -425}$   &  1129$^{\rm +1826}_{\rm -697}$  \\
188310	& G9.5 III  & 666$^{\rm +166}_{\rm -115}$    &  666$^{\rm +332}_{\rm -169}$  \\
219449	& K1 III    & 4548$^{\rm +691}_{\rm -2695}$  &  4548$^{\rm +1494}_{\rm -3188}$ \\
222404	& K1 III-IV & 3957$^{\rm +943}_{\rm -754}$   &  3957$^{\rm +2214}_{\rm -1249}$ \\
\enddata
\tablecomments{Spectral type is from SIMBAD, and the age is from the MIST model. The notation $1\sigma$ and $2\sigma$ indicate the confidence intervals that correspond to the uncertainties in age. \\}
\end{deluxetable}


\begin{deluxetable}{cccc}
\tablewidth{0pc}
\tabletypesize{\normalsize}
\tablecaption{Stellar Mass Comparison.\label{mass_table}}
\tablehead{
 \colhead{}   & \colhead{$M_{\rm EPE}$} & \colhead{$M_{\rm NPOI},1\sigma$} & \colhead{$M_{\rm NPOI},2\sigma$}   \\
 \colhead{HD} & \colhead{($M_\odot$)}   & \colhead{($M_\odot$)} & \colhead{($M_\odot$)}     }
\startdata
9826   & 1.27$\pm$0.06   & 1.23$^{\rm +0.02}_{\rm -0.04}$    & 1.23$^{\rm +0.03}_{\rm -0.09}$   \\ 
10700  & 0.783$\pm$0.012 & 0.876$^{\rm +0.006}_{\rm -0.007}$ & 0.876$^{\rm +0.013}_{\rm -0.014}$  \\
12929  & 1.5$\pm$0.2     & 1.029$\pm$0.010                   & 1.029$^{\rm +0.010}_{\rm -0.037}$    \\
20902  & 7.3$\pm$0.3     & 6.12$^{\rm +0.08}_{\rm -0.05}$    & 6.12$^{\rm +0.16}_{\rm -0.09}$    \\
28305  & 2.7$\pm$0.1     & 2.74$^{\rm +0.09}_{\rm -0.12}$    & 2.74$^{\rm +0.15}_{\rm -0.24}$  \\
54719  & 2.3$\pm$0.3     & 3.89$^{\rm +0.09}_{\rm -0.07}$    & 3.89$^{\rm +0.15}_{\rm -0.21}$  \\
62509  & 1.47$\pm$0.45   & 1.37$^{\rm +0.02}_{\rm -0.01}$    & 1.37$^{\rm +0.04}_{\rm -0.03}$  \\
66141  & 1.1$\pm$0.1     & 1.96$^{\rm +0.02}_{\rm -0.15}$   & 1.96$^{\rm +0.43}_{\rm -0.29}$  \\
120136 & 1.3             & 1.37$\pm$0.03                     & 1.37$^{\rm +0.05}_{\rm -0.06}$  \\
136726 & 1.8$\pm$0.25    & 2.60$^{\rm +0.20}_{\rm -0.25}$    & 2.60$^{\rm +0.48}_{\rm -0.44}$   \\
137759 & 1.4$\pm$0.2     & 1.95$^{\rm +0.21}_{\rm -0.11}$    & 1.95$^{\rm +0.44}_{\rm -0.20}$  \\
141004 & 1.275$\pm$0.089 & 1.09$^{\rm +0.06}_{\rm -0.05}$    & 1.09$^{\rm +0.11}_{\rm -0.10}$  \\
143107 & 1.7$\pm$0.1     & 1.69$^{\rm +0.29}_{\rm -0.30}$    & 1.69$^{\rm +0.58}_{\rm -0.55}$  \\
163917 & 3.04$\pm$0.06   & 3.098$^{\rm +0.017}_{\rm -0.016}$ & 3.098$^{\rm +0.031}_{\rm -0.035}$  \\
170693 & 0.98$\pm$0.05   & 1.88$^{\rm +0.36}_{\rm -0.33}$    & 1.88$^{\rm +0.68}_{\rm -0.56}$  \\
188310 & 2.2             & 2.40$\pm$0.12                     & 2.40$^{\rm +0.20}_{\rm -0.28}$   \\
219449 & 1.4$\pm$0.1     & 1.29$^{\rm +0.40}_{\rm -0.10}$    & 1.29$^{\rm +0.59}_{\rm -0.16}$  \\
222404 & 1.4$\pm$0.12    & 1.38$\pm$0.08                     & 1.38$^{\rm +0.15}_{\rm -0.16}$  \\
\enddata
\tablecomments{$M_{\rm EPE}$ is the stellar mass from \emph{EPE}, and $M_{\rm NPOI}$ is the stellar mass we get by using the $R$ and $T_{\rm eff}$ listed in Table \ref{sample} based on NPOI measurements. The notation $1\sigma$ and $2\sigma$ indicate the confidence intervals that correspond to the uncertainties in mass.
}
\end{deluxetable}

\clearpage


\startlongtable
\begin{deluxetable}{cccccc}
\tablewidth{0pc}
\tabletypesize{\normalsize}
\tablecaption{Planetary System Parameters.\label{planet_sys}}
\tablehead{
 \colhead{} & \colhead{} & \colhead{$P$} & \colhead{$K$} & \colhead{ } & \colhead{Source} \\
 \colhead{HD} & \colhead{} & \colhead{(d)} & \colhead{(m s$^{-1}$)} & \colhead{$e$} & \colhead{Notes}  }
\startdata
9826 	 & b & 4.61711$\pm$0.00018  & 70.519$\pm$0.0368  & 0.01186$\pm$0.006   &  1 \\
	 & c & 240.937$\pm$0.06     & 53.498$\pm$0.55   & 0.2445$\pm$0.1      &  1 \\
	 & d & 1281.439$\pm$2.0     & 67.7$\pm$0.55     & 0.316$\pm$0.07      &  1 \\
	 & e & 3848.86$\pm$0.74     &                   & 0.00536$\pm$0.00044 &  1 \\
10700	 & e & 162.87$\pm$1.08      & 0.55$\pm$0.13     & 0.18$\pm$0.18       &  1 \\
	 & f & 636.13$\pm$47.69     & 0.35$\pm$0.12     & 0.16$\pm$0.16       &  1 \\
	 & g & 20.0$\pm$0.2         & 0.49$\pm$0.11     & 0.06$\pm$0.13       &  1 \\
12929	 & b & 380.8$\pm$0.3        & 41.1$\pm$0.8      & 0.25$\pm$0.03       &  2 \\
20902	 & b & 128.2$\pm$0.1        & 70.8$\pm$1.5      & 0.1$\pm$0.04        &  3 \\
28305	 & b & 594.9$\pm$5.3        & 93.2$\pm$2.1      & 0.151$\pm$0.023     &  1 \\
54719	 & b & 305.5$\pm$0.1        & 350.5$\pm$3.4     & 0.031$\pm$0.009     &  1 \\
62509	 & b & 589.64$\pm$0.81      & 46.0$\pm$1.6      & 0.02$\pm$0.03       &  1 \\
66141	 & b & 480.5$\pm$0.5        & 146.2$\pm$2.7     & 0.07$\pm$0.03       &  4 \\
120136	 & b & 3.31249$\pm$3.13E-05 & 469.59$\pm$14.86  & 0.0787$\pm$0.0382   &  1 \\
136726	 & b & 516.22$\pm$3.25      & 189.7$\pm$7.15    & 0.08$\pm$0.03       &  1 \\
137759	 & b & 510.855$\pm$0.014    & 311.0$\pm$1.0     & 0.7008$\pm$0.0018   &  1 \\
141004	 & b & 15.5083$\pm$0.0018   & 3.52$\pm$0.36     & 0.16$\pm$0.11       &  5 \\
143107	 & b & 417.9$\pm$0.5        & 129.4$\pm$2.0     & 0.11$\pm$0.03       &  6 \\
163917	 & b & 530.21$\pm$0.1       & 288.26$\pm$1.0    & 0.124$\pm$0.003     &  1 \\
	 & c & 3184.83$\pm$5.93     & 177.78$\pm$1.3    & 0.18$\pm$0.006      &  1 \\
170693	 & b & 479.1$\pm$6.2        & 110.5$\pm$7.0     & 0.38$\pm$0.06       &  7 \\
188310	 & b & 136.75$\pm$0.25      & 65.4$\pm$1.7      & N/A                 &  8 \\
219449	 & b & 181.4$\pm$0.1        & 91.0$\pm$2.3      & 0.027$\pm$0.026     &  1 \\
222404	 & b & 905.64$\pm$2.83      & 28.08$\pm$1.45    & 0.0724$\pm$0.0679   &  1 \\
\enddata
\tablecomments{ 1. All parameters from \emph{EPE}, and when there were asymmetric error bars, we used the larger one here. The number of significant figures is erratic throughout the table, but it is how the values were presented in the catalog. We started with \emph{EPE} for planetary orbital parameters, but not all stars had $K$ values in \emph{EPE}. In those cases, we used $K$ from:} 2. \citet{2011AandA...529A.134L}; 3. \citet{2012AandA...543A..37L}; 4. \citet{2012AandA...548A.118L}; 5. \citet{2021ApJS..255....8R}; 6. \citet{2012AandA...546A...5L}; 7. \citet{2009AandA...499..935D}; and 8. \citet{2008PASJ...60..539S}.  
\end{deluxetable}

\clearpage


\begin{deluxetable}{lccc}
\tablewidth{0pc}
\tabletypesize{\normalsize}
\tablecaption{Planetary Mass Comparison.\label{planet_mass_table}}
\tablehead{
 \colhead{} & \colhead{\emph{EPE}} & \colhead{Initial} & \colhead{Final}   \\
 \cline{2-4} 
 \colhead{} & \colhead{$m \sin i$}      & \colhead{$m \sin i$} & \colhead{$m \sin i$}  \\
 \colhead{HD} & \colhead{($M_{\rm Jup}$)} & \colhead{($M_{\rm Jup}$)} & \colhead{($M_{\rm Jup}$)} }
\startdata
9826	b & 0.62$\pm$0.09      & 0.66   & 0.66$\pm$0.02 \\
9826	c & 1.8$\pm$0.26       & 1.82   & 1.82$\pm$0.05 \\
9826	d & 10.19              & 3.94   & 3.95$\pm$0.11 \\
10700	e & 0.0124$\pm$0.0026  & 0.013  & 0.013$\pm$0.003 \\
10700	f & 0.0124$\pm$0.0043  & 0.011  & 0.011$\pm$0.004 \\
10700	g & 0.00551$\pm$0.0013 & 0.005  & 0.005$\pm$0.001 \\
12929	b & 1.8$\pm$0.2        & 1.45   & 1.45$\pm$0.03 \\
20902	b & 6.6$\pm$0.2        & 5.84   & 5.85$\pm$0.14 \\
28305	b & 7.34$\pm$0.05      & 7.46   & 7.47$\pm$0.32 \\
54719	b & 20.6               & 28.69  & 28.83$\pm$0.47 \\
62509	b & 2.9$\pm$0.1        & 2.34   & 2.34$\pm$0.09 \\
66141	b & 6.0$\pm$0.3        & 8.80   & 8.82$\pm$0.18 \\
120136	b & 4.13$\pm$0.35      & 4.23   & 4.24$\pm$0.15 \\
136726	b & 11.0873$\pm$1.1    & 14.10  & 14.15$\pm$1.23 \\
137759	b & 16.4$\pm$9.3       & 13.61  & 13.67$\pm$0.44  \\
141004	b & 0.0428$\pm$0.0047  & 0.05   & 0.05$\pm$0.01 \\
143107	b & 6.7$\pm$0.3        & 6.71   & 6.72$\pm$0.98 \\
163917	b & 22.206             & 24.19  & 24.31$\pm$0.14 \\
170693	b & 3.88$\pm$0.85      & 5.99   & 6.00$\pm$1.01 \\
188310	b & 2.02$\pm$0.2       & 2.97   & 2.97$\pm$0.14 \\
219449	b & 3.2                & 3.00   & 3.00$\pm$0.76 \\
222404	b & 1.742$\pm$0.075    & 1.65   & 1.65$\pm$0.12 \\
\enddata
\tablecomments{The columns show $m \sin i$ from \emph{EPE}, and ``Initial'' and ``Final'' are the values for the first and second iteration, respectively, as described in Section 4. We used the larger of the error bars for the $1\sigma$ CI as listed in Table \ref{mass_table}.}
\end{deluxetable}

\clearpage


\begin{deluxetable}{cccccc}
\tablewidth{0pc}
\tabletypesize{\normalsize}
\tablecaption{Habitable Zone Extent.\label{hab_zone}}
\tablehead{
 \colhead{} & \colhead{$T_{\rm eff}$} & \colhead{$L$} & \colhead{$r_i$} & \colhead{$r_o$} & \colhead{$a$}  \\
 \colhead{HD} & \colhead{(K)} & \colhead{($L_\odot$)} & \colhead{(AU)} & \colhead{(AU)} & \colhead{(AU)}   }
\startdata
9826	b & 6114$\pm$77  & 3.1$\pm$0.1     & 1.43$\pm$0.05  & 2.84$\pm$0.08  & 0.059$\pm$0.001  \\
9826	c &              &                 &                &                & 0.861  \\
9826	d &              &                 &                &                & 2.55   \\
10700	e & 5301$\pm$13  & 0.471$\pm$0.001 & 0.60$\pm$0.002 & 1.18$\pm$0.002 & 0.538$\pm$0.006  \\
10700	f &              &                 &                &                & 1.334$\pm$0.044  \\
10700	g &              &                 &                &                & 0.133$\pm$0.002  \\
12929	b & 4373$\pm$8   & 76.2$\pm$0.8    & 8.20$\pm$0.06  & 16.07$\pm$0.11 & 1.2    \\
20902	b & 5859$\pm$41  & 2994$\pm$178    & 45.39$\pm$1.94 & 90.04$\pm$3.78 & 0.97   \\
28305	b & 4880$\pm$7   & 79.4$\pm$3.4    & 8.06$\pm$0.24  & 15.82$\pm$0.47 & 1.9$\pm$0.07    \\
54719	b & 4583$\pm$70  & 365$\pm$14      & 17.68$\pm$0.51 & 34.67$\pm$0.97 & 1.17   \\
62509	b & 4586$\pm$57  & 32.7$\pm$0.3    & 5.29$\pm$0.05  & 10.38$\pm$0.06 & 1.69$\pm$0.03   \\
66141	b & 4521$\pm$65  & 209$\pm$9       & 13.45$\pm$0.41 & 26.38$\pm$0.77 & 1.2$\pm$0.1    \\
120136	b & 6556$\pm$212 & 3.2$\pm$0.1     & 1.38$\pm$0.07  & 2.77$\pm$0.04  & 0.046  \\
136726	b & 4358$\pm$59  & 276$\pm$14      & 15.63$\pm$0.57 & 30.65$\pm$1.10 & 1.54$\pm$0.07   \\
137759	b & 4756$\pm$29  & 63.9$\pm$1.6    & 7.30$\pm$0.13  & 14.32$\pm$0.26 & 1.448$\pm$0.029  \\
141004	b & 6087$\pm$174 & 1.96$\pm$0.01   & 1.14$\pm$0.04  & 2.26$\pm$0.02  & 0.1238$\pm$0.002 \\
143107	b & 4408$\pm$109 & 194$\pm$5       & 13.05$\pm$0.27 & 25.60$\pm$0.43 & 1.3    \\
163917	b & 5000$\pm$63  & 96.6$\pm$2.1    & 8.81$\pm$0.16  & 17.30$\pm$0.26 & 1.79   \\
170693	b & 4714$\pm$101 & 177$\pm$15      & 12.18$\pm$0.76 & 23.89$\pm$1.47 & 1.19$\pm$0.01   \\
188310	b & 5263$\pm$105 & 71.7$\pm$5.3    & 7.42$\pm$0.41  & 14.60$\pm$0.77 & 0.58$\pm$0.03   \\
219449	b & 4730$\pm$68  & 52.7$\pm$1.2    & 6.64$\pm$0.12  & 13.03$\pm$0.20 & 0.7    \\
222404	b & 4792$\pm$62  & 11.1$\pm$0.3    & 3.03$\pm$0.07  & 5.95$\pm$0.12  & 2.1459$\pm$0.0048 \\
\enddata
\tablecomments{$T_{\rm eff}$ and $L$ are from the papers listed in column 11 of Table \ref{sample} ($L_\odot = 3.826 \times 10^{\rm 33}$), $r_o$ and $r_i$ are the inner and outer boundaries of the habitable zone, respectively, as calculated in Section 4, and $a$ is the planet's distance from the star (\emph{EPE}).}
\end{deluxetable}


\clearpage

\clearpage

\begin{figure}[!h]
   \centering
   \includegraphics[width=8cm]{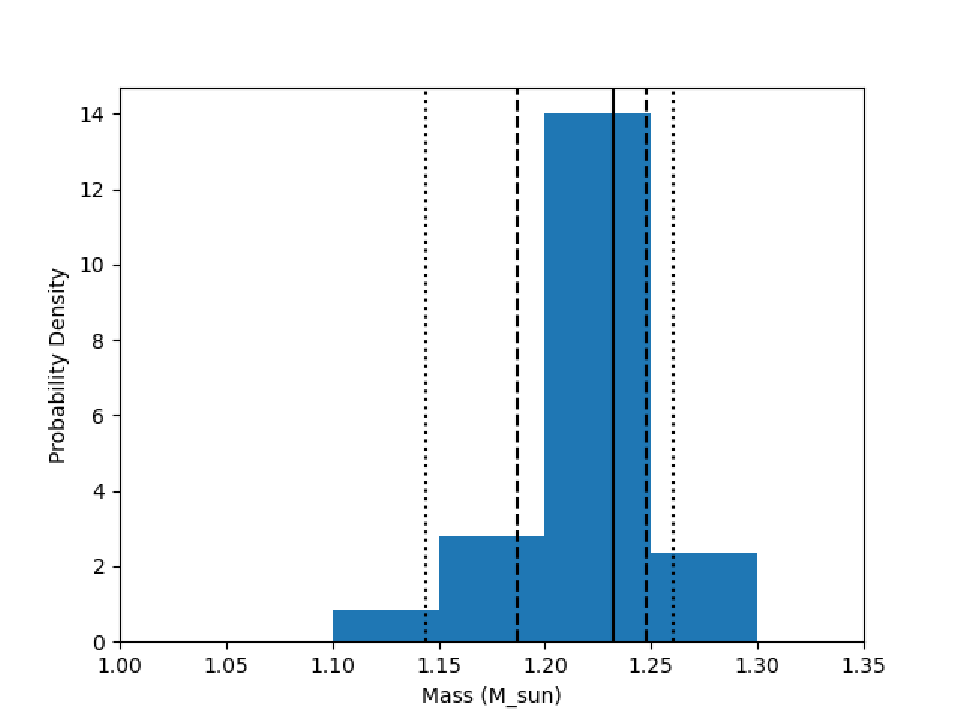} 
   \includegraphics[width=8cm]{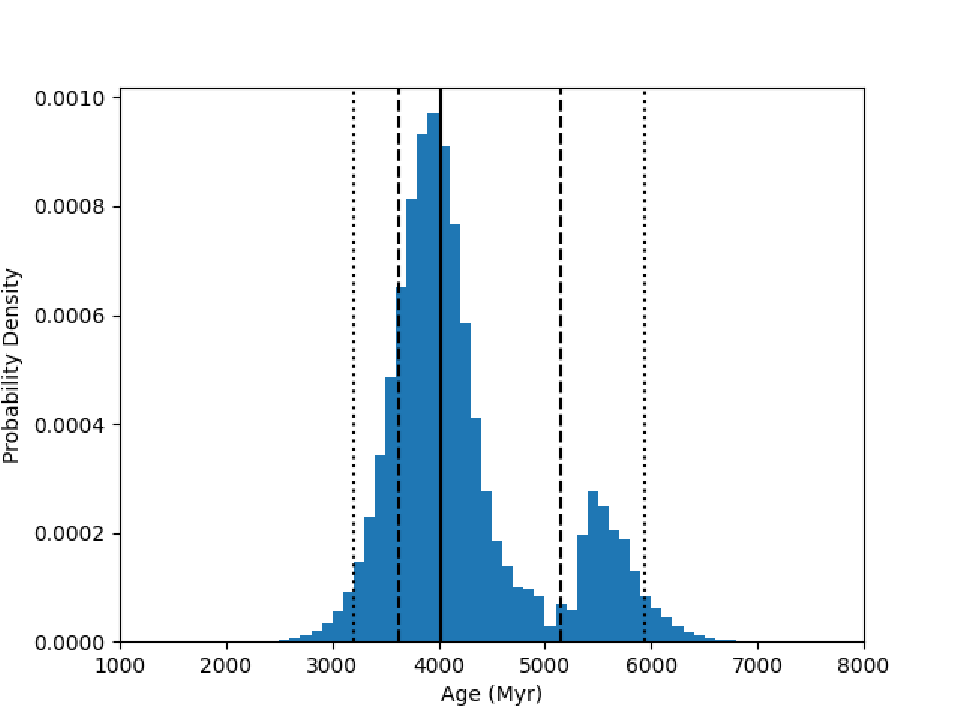}
   \includegraphics[width=8cm]{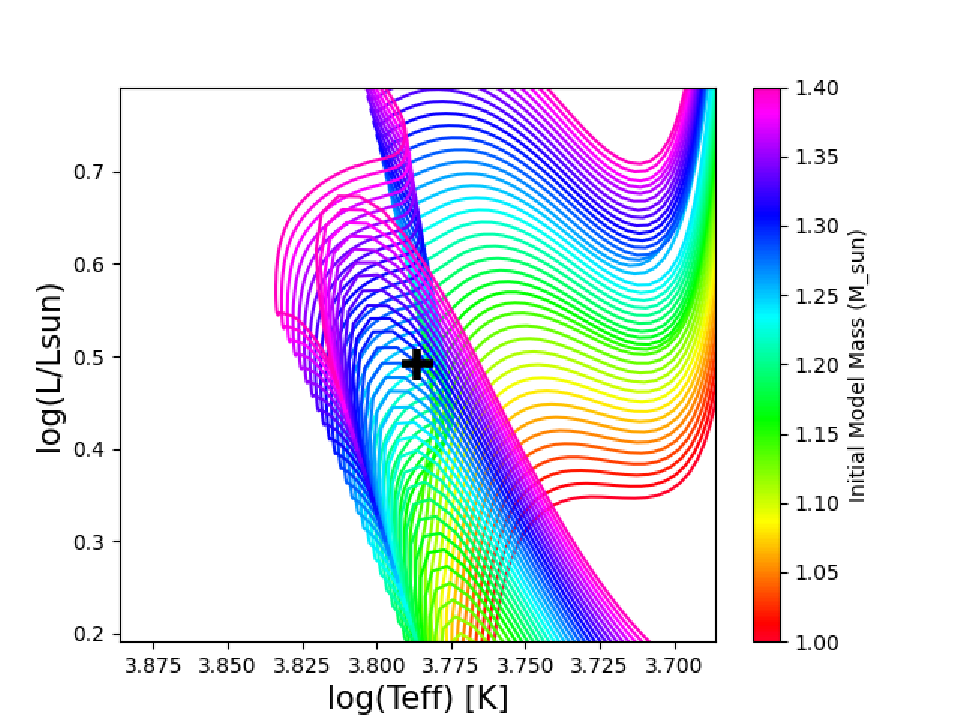}
   \includegraphics[width=8cm]{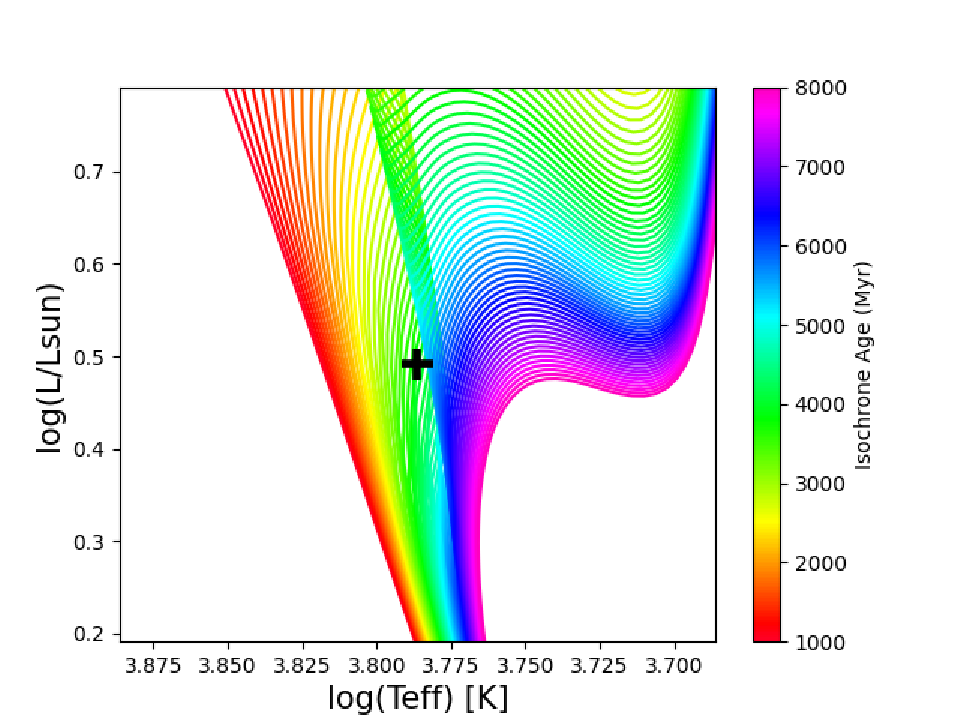}
   \includegraphics[width=8cm]{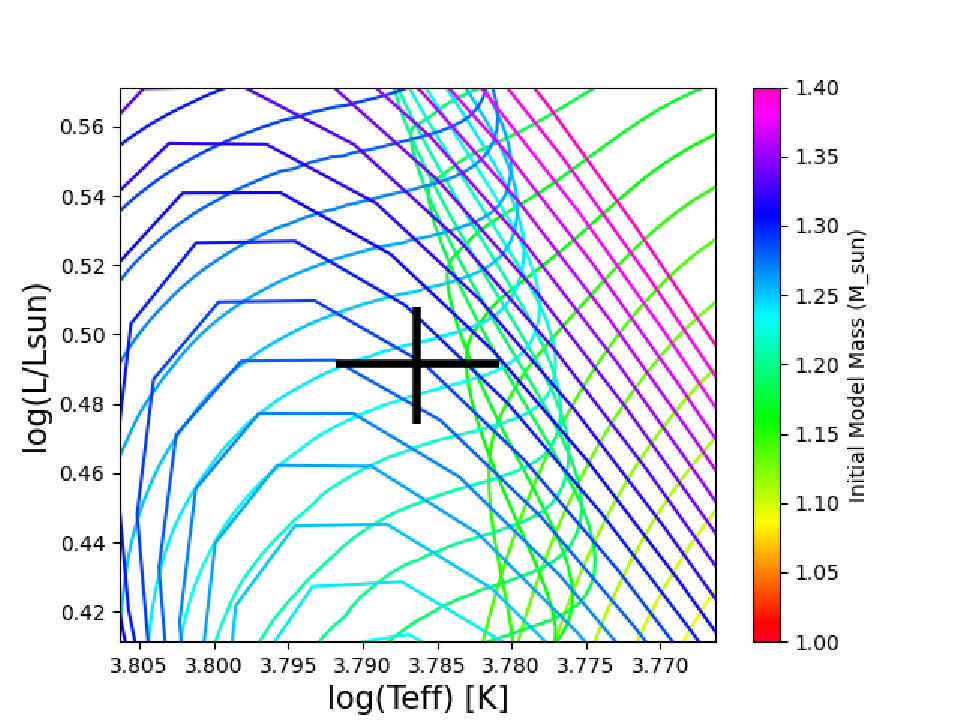}
   \includegraphics[width=8cm]{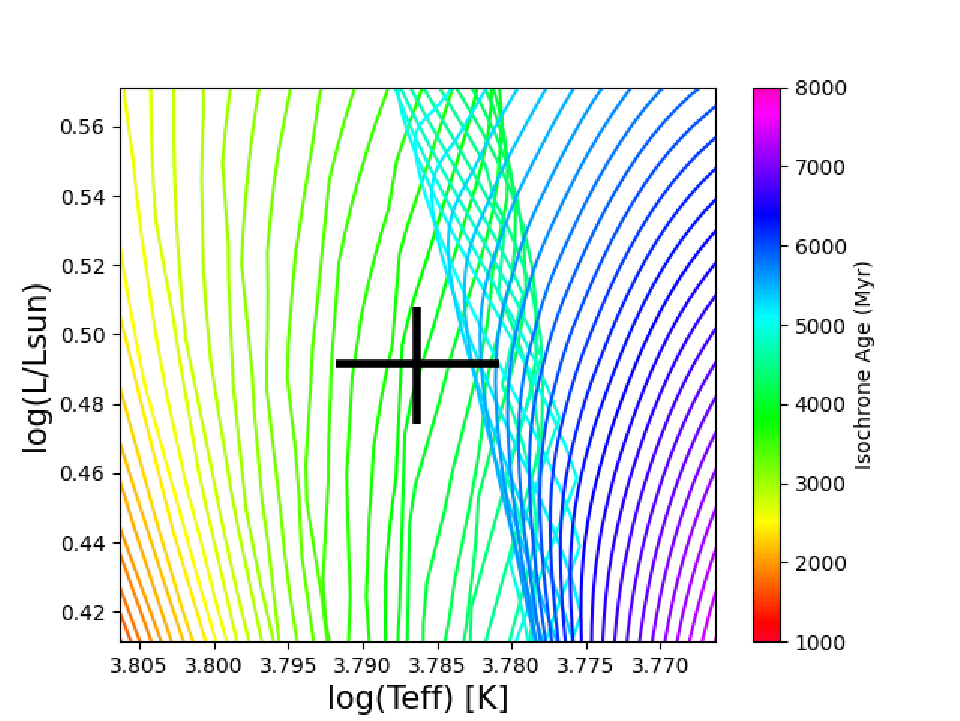}
\caption{The results from our age (\emph{left}) and mass (\emph{right}) fits for HD 9826/$\upsilon$ And. Note that because we used the mass tracks to calculate age and the isochrones to calculate mass, this may seem counter-intuitive to the way the plots are labelled (see Section 3.1 for discussion on this process). The top row shows the PDF results where the solid black vertical line shows the median, the dashed lines show the 1-$\sigma$ CI, and the dotted lines show the 2-$\sigma$ CIs. The second row shows the HR diagram with HD 9826's data point and error bars as the black cross, and the bottom row shows a zoomed-in version of the second row. The numerical values are in Tables \ref{mass_table} and \ref{ages}. The rest of the plots are available via the online version of \emph{The Astrophysical Journal}.}
\end{figure}
   
\clearpage

\begin{figure}[h]
\includegraphics[width=1.0\textwidth]{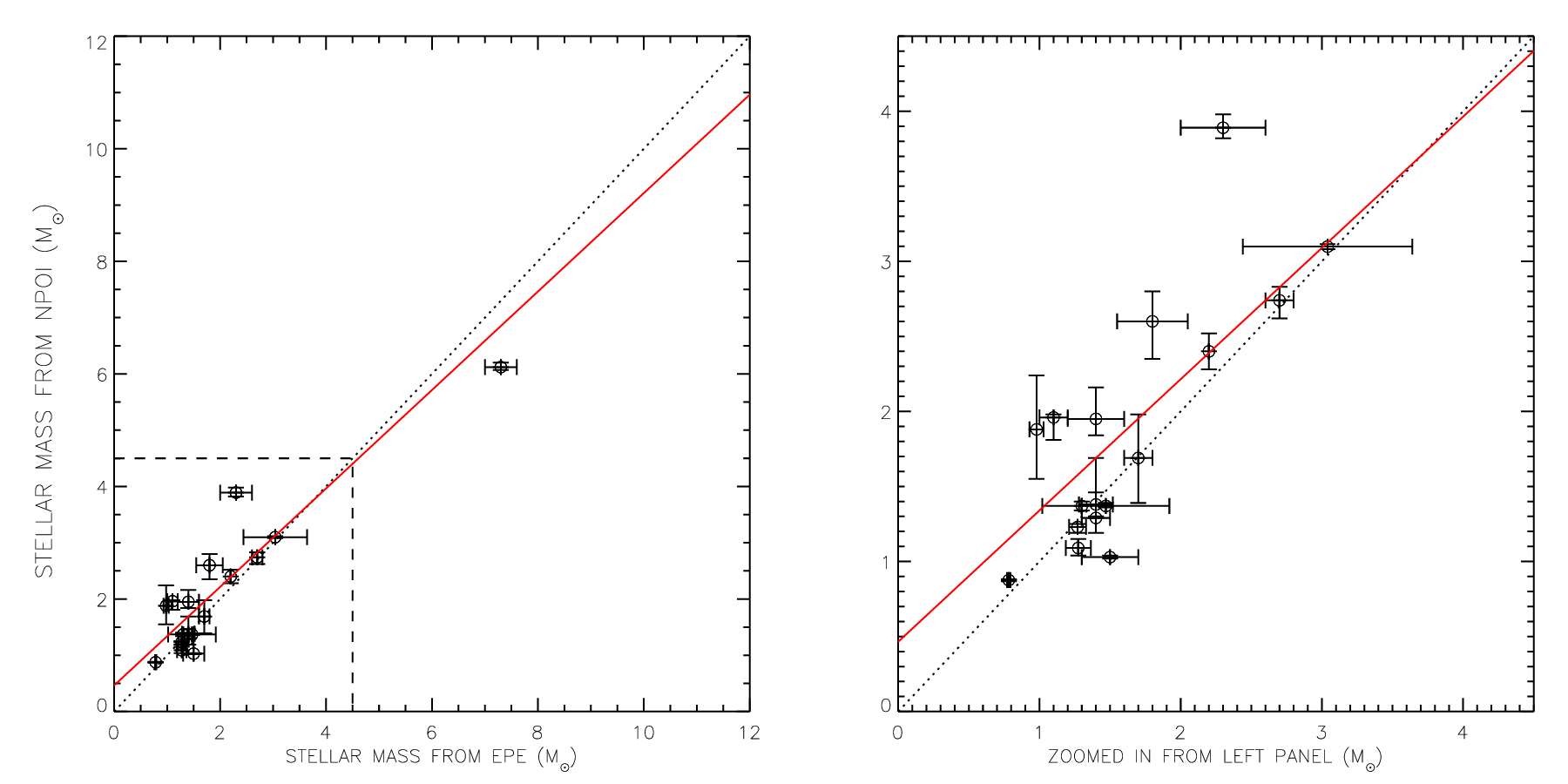}
\caption{Comparing the stellar mass from \emph{EPE} and those based on NPOI measurements. All the stars are shown in the \emph{left panel}, while a zoomed-in version is shown in the \emph{right panel} so there is not as much crowding. The dotted line is the 1:1 ratio, the dashed lines in the left panel show which section of the plot is magnified in the right panel, and the solid red line is the linear fit of $f(x) = 0.788 x - 0.624$.
}
  \label{star_mass}
\end{figure}

\clearpage

\begin{figure}[h]
\includegraphics[width=0.85\textwidth]{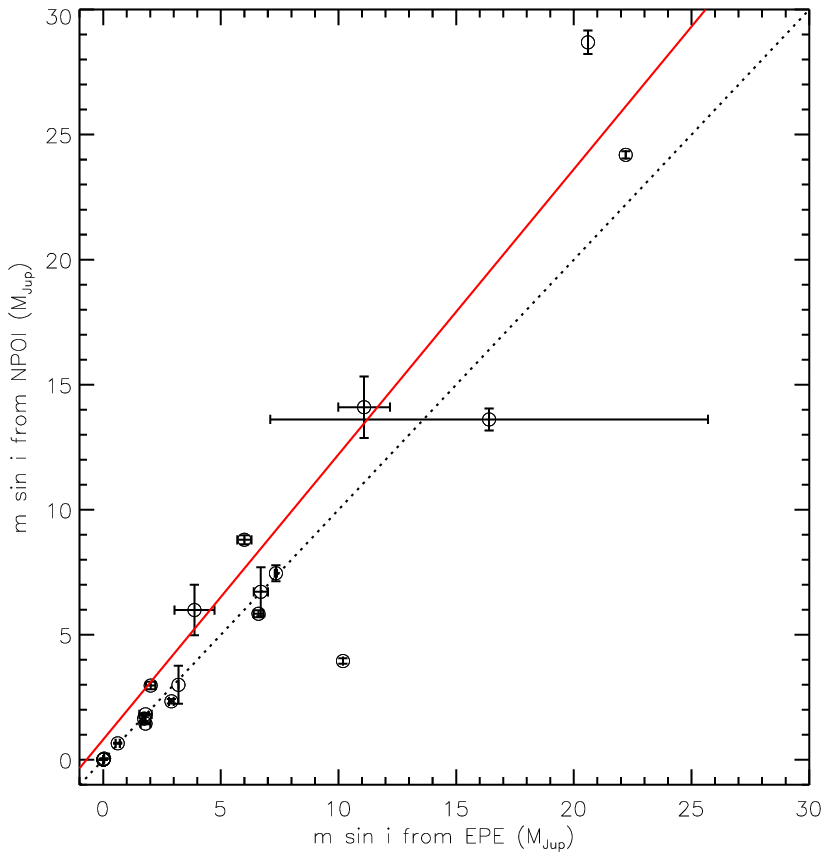}
\caption{Comparing $m \sin i$ from \emph{EPE} and $m \sin i$ based on NPOI measurements. The dotted line is the 1:1 ratio, and the solid red line is the linear fit of $f(x) = 1.124 x - 0.335$.}
  \label{planet_mass}
\end{figure}

\clearpage

\end{document}